\documentstyle[twoside,fleqn,espcrc2]{article}

\newcommand{\AmS}{{\protect\the\textfont2
  A\kern-.1667em\lower.5ex\hbox{M}\kern-.125emS}}

\newcommand{\beq}{\begin{equation}}
\newcommand{\eeq}{\end{equation}}
\newcommand{\bea}{\begin{eqnarray}}
\newcommand{\eea}{\end{eqnarray}}
\newcommand{\als}{\alpha_s}
\newcommand{\aq}{a(q^2)}

\title{Global approximants with renormalization scale 
invariance in pQCD}

\author{G. Cveti\v c\address{Department of Physics, University 
Dortmund, D-44221 Dortmund}%
\thanks{hep-ph/9808273, based on an invited talk at QCD'98, M
ontpellier, France, July 1998}}

\begin{document}

\begin{abstract}
Truncated perturbative series (TPS's) of any observable have the
unphysical dependence on the choice of the renormalization scale
(RScl). The diagonal Pad\'e approximants (dPA's) to any TPS of
an observable possess the favorable property
of being invariant in the large-${\beta_0}$ limit. This means that
they are invariant under the change of the RScl $\mu^2$ when the 
``running'' coupling parameter ${\alpha}_s(\mu^2)$ evolves 
according to the one-loop renormalization group equation. 
We present a method which generalizes this result -- the 
resulting new approximants are fully RScl-invariant in the
perturbative QCD (pQCD). Further, we present some numerical 
examples. Both the dPA's and the new approximants are global, 
i.e., their structure goes beyond the usual (polynomial) TPS form 
and thus they could reveal some non-perturbative effects.
\end{abstract}

\maketitle

The contribution is based partly on \cite{ours}. It
contains additionally some numerical examples. 
Since the construction of our approximants is related
with the diagonal Pad\'e approximants (dPA's), we first 
explain what PA's are.

\section{What is Pad\'e approximant ?}

Suppose we have a physical quantity $F(z)$ which depends
on the parameter $z$. In the
expansion of $F(z)$ in powers of $z$
\beq
F(z) = f_0 + f_1 z + \cdots f_n z^n + \cdots,
\label{Fz}
\eeq

\vspace{-0.1cm}

\noindent the $f_j$'s are in principle calculable.
Suppose that $f_0,\!\ldots,\!f_{L+M}$ have been calculated, i.e., 
the truncated perturbation series (TPS) $F_{[L+M]}(z)$ is known.
Then, the PA $[L/M]_F(z)$ of order $L/M$ to $F(z)$ is defined 
as the ratio of two polynomials, of degree $L$ (nominator)
and $M$ (denominator), such that, when expanded back in powers
of $z$, it reproduces the first $L+M+1$ terms of (\ref{Fz}).
In general, this condition determines uniquely the PA. 
It is the minimal condition that any approximant 
to the TPS $F_{[L+M]}(z)$ has to fulfill. PA has, 
in addition, the favorable property that 
it goes beyond the analytic form, showing structures
(pole singularities in the complex $z$ plane) 
not explicitly contained in series (\ref{Fz}). 
We call such approximants {\em global\/}, since 
they can give us clues to some nonperturbative 
properties of $F(z)$ at large $|z|$. 
The diagonal PA's (dPA's) are those with $L\!=\!M$.

\section{Construction of new approximants}

A generic observable $S$ in pQCD is
\bea
\lefteqn{
S \equiv a(q^2) f(q^2) = a(q^2) [
1 + }
\nonumber\\
&& r_1(q^2) a(q^2) + \cdots 
+ r_n(q^2) a^n(q^2) + \cdots ] ,
\label{S}
\eea
where $q^2$ is a chosen renormalization scale (RScl),
$\aq\!\equiv\!\als(q^2)/\pi$. $S$ is RScl-independent.
However, available is only a TPS $S_{[n]}$
\bea
\lefteqn{
S_{[n]}(q^2) \equiv a(q^2) f^{[n]}(q^2) = a(q^2) [
1 + }
\nonumber\\
&&r_1(q^2) a(q^2) + \cdots 
+ r_n(q^2) a^n(q^2) ].
\label{Sn}
\eea
It has unphysical RScl-dependence (truncation).

How to extract as much information as possible from
the available TPS (\ref{Sn})? Firstly, any approximant should
fulfill the minimal condition (cf.~previous Sec.).
Secondly, the full observable $S$
contains, in general, non-perturbative effects not
explicitly manifested in power series (\ref{S}) -- this
leads us to consider {\em global\/} approximants 
(cf.~previous Sec.). Thirdly, $S$ is RScl-independent,
so it is natural to expect that RScl-invariant approximants
bring us closer to $S$.
So, the question here is: How to construct {\em global\/}
approximants which are based on the TPS $S_{[n]}(q^2)$ and
are RScl-independent?

It turns out that a partial answer to this question is the
diagonal Pad\'e approximant (dPA) $[M/M]_S(a)$ \cite{Gardi}.
To see this, we recall that the evolution of 
$a(p^2)$ with the change of RScl $p^2$ is governed 
in pQCD by the RGE
\beq
\frac{d a}{d \ln (-p^2)} = 
- \beta_0 a^2 ( 1 + c_1 a + c_2 a^2 + \cdots ).
\label{RGE}
\eeq
In the large-$\beta_0$ limit (i.e., $c_1=c_2=\cdots=0$)
\beq
a^{(1l.)}(p^2) = \aq \left[1 + \ln(p^2/q^2) \beta_0 \aq 
\right]^{-1} .
\label{1lsol}
\eeq
Thus, the RScl change $q^2\!\mapsto\!p^2$
results in the change of $z\!\equiv\!\aq$:
$z\!\mapsto\!z/(1\!+\!b z)$.
The dPA $[M/M]_S(z)$ {\em is\/} invariant under 
this argument transformation, i.e., RScl-invariant
when $a(p^2)$ evolves according to the one-loop RGE.
$[M/M]_S(z)$ reproduces the TPS $S_{[2M-1]}$
(\ref{Sn}) to all the available powers (cf.~also
the previous Sec.).

Is it possible to go beyond this large-$\beta_0$ approximation?
The question was raised first by the authors of 
\cite{Brodskyetal}. The answer: it is. 

The dPA approach has to be extended. We describe the
algorithm leading to the new, fully
RScl-invariant, approximants to $S$. Introduce
\bea
k_m \left( a(q^2) \right)& \equiv& \frac{1}{\aq}
\frac{1}{m!} \frac{d^m a({\tilde p}^2)}{d (\ln
{\tilde p}^2 )^m} {\Big |}_{{\tilde p}^2=q^2}\ ;
\label{km}
\\
k_m(a)& =& (-1)^m \beta_0^m a^m [ 1 + {\cal O} (c_1 a) ] \ ,
\label{kjs}
\eea
where higher orders in (\ref{kjs}) 
can be calculated by RGE (\ref{RGE}).
We rearrange power series (\ref{S}) for $S$ 
into perturbation series in $k_m$ ($\sim$$a^m$)
\bea
\lefteqn{
S \equiv a(q^2) f(q^2) = \aq [ 1 + f_1(q^2) k_1(a(q^2)) +}
\nonumber\\
&& + \cdots + f_n(q^2) k_n(a(q^2)) +
\cdots ],
\label{Snew}
\eea
where $f_n$ is uniquely determined by the
first $n$ coefficients $r_j$ of (\ref{S}).
For example, $f_1$$=$$-r_1/\beta_0$, 
$f_2$$=$$(-c_1 r_1\!+\!r_2)/\beta_0^2$,
etc. Thus, knowing the TPS $S_{[2M-1]}(q^2)$
($r_1,\!\ldots,r_{2M-1}$) in powers of $a$, 
we know the corresponding TPS in $k_m$'s of 
(\ref{Snew}) up to (and including) the
$f_{2M-1}$-term.
Define the (power) TPS, by the large-$\beta_0$ substitution
[cf.~(\ref{kjs})]
$k_m\!\mapsto\!(-1)^m \beta_0^m a(q^2)^m$
\bea
\lefteqn{
\aq {\cal F}^{[2M-1]}(q^2) = a [1 - f_1 \beta_0 a + }
\nonumber\\
&&+\cdots + f_{2M-1} (-1)^{2M-1} \beta_0^{2M-1} a^{2M-1} ].
\label{TPSfPA}
\eea
The dPA $[M/M](a)$ for this (power) TPS
can be decomposed into a sum of simple fractions
\beq
[M/M]_{a {\cal F}}(a(q^2)) = 
 \sum_{j=1}^M {\tilde \alpha}_j
\frac{\aq}{\left[1 + \left({\tilde u}_j \beta_0 \right) \aq \right]}.
\label{dPAaF2}
\eeq
We now denote
\beq
p_j^2 = q^2 \exp[{\tilde u}_j(q^2)], \ {\rm i.e.} \ 
{\tilde u}_j(q^2) = \ln(p_j^2/q^2),
\label{pjdef}
\eeq
and therefore, in view of (\ref{1lsol}),
rewrite (\ref{dPAaF2}) as
\beq
[M/M]_{a {\cal F}} = \sum_{j=1}^M {\tilde \alpha}_j
a^{(1l.)}(p_j^2) \ .
\label{dPAaF3}
\eeq

\vspace{-0.1cm}

\noindent Replace here 
$a^{(1l.)}(p_j^2)\!\mapsto\!a(p_j^2)$, 
where $a(p_j^2)$ is evolved
from $a(q^2)$ via the full RGE (\ref{RGE})
\beq
A_S^{[M/M]} = \sum_{j=1}^M {\tilde \alpha}_j a(p_j^2) \ .
\label{result}
\eeq

\vspace{-0.1cm}

\noindent This approximant satisfies all
the conditions that we set forth at the outset:
a) when expanded back in powers of $a(q^2)$
($q^2$ is the original RScl), it reproduces
all the terms of the TPS $S_{[2M-1]}(q^2)$ (\ref{Sn});
b) it is global since it is a modification
of the dPA approach (the latter is global);
c) it is fully RScl-invariant in the pQCD-sense, 
i.e., independent of the initial choice of the RScl $q^2$,
where $a(q^2)$ evolves according to the full available
perturbative RGE (\ref{RGE}).
Explicit proof is given in \cite{ours} (first entry). 
In fact, coefficients
${\tilde \alpha}_j$ and squared momenta $p_j^2$
are all RScl-invariant.

Approximants (\ref{result}) can be applied directly
only to TPS's $S_{[n]}$ (\ref{Sn}) 
with an odd number $n=2M\!-\!1$.
In pQCD, $S_{[1]}$ are available for many observables,
$S_{[2]}$ for a few, $S_{[3]}$ for none.
Thus, (\ref{result}) is applicable only to
TPS's $S_{[1]}$. In this case ($M\!=\!1$), 
(\ref{result}) reduces to the 
effective charge (cf.~\cite{Grunberg}).
To apply (\ref{result}) to 
$S_{[2]}(q^2)$, a modification is needed. 
Introduce a new observable ${\tilde S}
\equiv S*S$
\bea
\lefteqn{
{\tilde S} = (S)^2 = \aq F(q^2) = \aq [0 + 1 \aq +} 
\nonumber\\
&&+ R_2(q^2) a^2(q^2) + R_3(q^2) a^3(q^2)+\cdots ],
\label{Stil}
\eea
where $R_1\!=\!1$, $R_2\!=\!2 r_1$, $R_3\!=\!r_1^2\!+\!2 r_2$, etc.
Having $S_{[2]}(q^2)$ (\ref{Sn}), we have 
${\tilde S}_{[3]}(q^2)$ (i.e., up to and including the
$R_3$-term).
We apply (\ref{result})
to ${\tilde S}_{3}$ ($M\!=\!2$).
Although the leading order term in (\ref{Stil}) is
$0*\aq$, the algorithm survives, results in
\beq
{\tilde S} = A^{[2/2]}_{\tilde S}\!+\!{\cal O} (a^5),
\ \ 
S = \sqrt{A^{[2/2]}_{\tilde S}}\!+\!{\cal O}(a^4) .
\label{Stilres}
\eeq
\beq
\sqrt{A^{[2/2]}_{\tilde S}}=\sqrt{ {\tilde \alpha}_1
\left[ a(p_1^2) - a(p_2^2) \right] },
\label{appexpl1}
\eeq
where ${\tilde \alpha}_1$ and $p^2_j$ are determined
by $r_1, r_2$ and ${\beta}_0, c_1, c_2$ (cf. \cite{ours}, 
last entry). $ \ $ Approximant (\ref{appexpl1})  

\newpage

\noindent is fully RScl-invariant.
The scales $p_j^2$ may be complex
in some cases, but the result is real.

\section{Specific numerical examples}

\subsection{A case of a Euclidean observable}

Consider the Bjorken polarized sum rule
\bea
\lefteqn{
\int_0^1 dx \left[ g_1^{(p)}(x,Q^2) -  g_1^{(n)}(x,Q^2)
\right] = }
\nonumber\\
&& [ |g_A|/(3 |g_V|)] [1 - S(-Q^2)],
\label{Bjpsr1}
\eea
$p^2\!=\!-Q^2$$<0$ is $\gamma^{\ast}$ 
momentum transfer. TPS $S_{[2]}(-Q^2)$ is now known: 
$r_1\!=\!3.584$ (\cite{GLZN}); $r_2\!=\!20.212$ (\cite{LV}).
These $r_j$'s are in the ${\overline {\rm MS}}$ scheme, 
$n_f\!=\!3$, and the chosen RScl $q^2_{\rm RScl}\!=\!-Q^2$.
We take $\sqrt{Q^2}\!=\!1.76$ GeV [$a(-Q^2)\!=\!0.083$].

First we note that any approximant
can be used also to predict the next
coefficient $r_3$, by reexpanding the approximant back 
in powers of $a$. We give in Table 1 results of 
various approximants to $S$ of (\ref{Bjpsr1}):
PA's, approximant (\ref{appexpl1}),
and the effective charge method (ECH \cite{Grunberg},
we set $c_3^{\rm (ECH)}\!=\!0$). 
In brackets, results are given for another scheme
($c_2\!=\!3$, $c_3\!=\!0$).
We used the known $c_3$ parameter of the ${\overline {\rm MS}}$
scheme \cite{QCDbeta} wherever possible, i.e.,
${\overline {\rm MS}}$ is characterized 
by $c_2\!=\!4.471$ and $c_3\!=\!20.99$.
\begin{table*}[hbt]
\setlength{\tabcolsep}{1.5pc}
\newlength{\digitwidth} \settowidth{\digitwidth}{\rm 0}
\catcode`?=\active \def?{\kern\digitwidth}
\caption{Predictions of various approximants for the
Bjorken polarized sum rule (with $n_f\!=\!3$) 
in the ${\overline {\rm MS}}$ ($c_2\!=\!4.471$) 
and the $c_2\!=\!3$ scheme (in brackets). 
$\sqrt{Q^2}\!=\!1.76$ GeV; RScl chosen: $q^2_{\rm RScl}\!=\!-Q^2$;
$a(-Q^2)\!=\!0.083$.}
\label{tab1}
\begin{tabular*}{\textwidth}{@{}l@{\extracolsep{\fill}}rrr}
\hline 
Approx. & $S(-Q^2)^{\rm pr.}$ & 
$r_3^{\rm pr.}$ & $ S_{[3]}( -Q^2; q^2_{\rm RScl} )^{\rm pr.}$
\\ 
\hline
\hline
$S_{[2]}$ (TPS) & $0.1192$ ($0.1175$) & --  & $0.1192$ ($0.1175$) \\
\hline
$[1/2]_S$ & $0.1273$ ($0.1261$) & $98.8$ ($109.4$) & 
$0.1239$ ($0.1223$) \\
$\sqrt{[2/2]_{S^2}}$ & $0.1326$ ($0.1322$) &
$125.2$ ($141.1$) & $0.1252$ ($0.1238$) \\
$\sqrt{A^{[2/2]}_{S^2}}$ & $0.1378$ ($0.1370$) &
$138.3$ ($151.4$) & $0.1258$ ($0.1242$) \\
\hline
$S_{[2]}^{({\rm ECH})}$ & $0.1320$ ($0.1320$) &
$129.9$ ($140.4$) & $0.1254$ ($0.1237$) \\
\hline
\end{tabular*}
\end{table*}
The results of method (\ref{appexpl1})
differ somewhat from those of other methods.
Further, values of $S(-Q^2)$
(cf.~second column) predicted by global methods
show up some scheme ($c_2-$)dependence ($\sim\!1\%$). 
The RScl-dependence is another
source of ambiguity in the results of the PA's. 
For example, if changing
RScl $Q\!\mapsto\!2 Q$, the result of the $[1/2]$ PA 
changes by $3.5\%$, that of $[2/2]^{1/2}$ by
$2 \%$, and the original TPS $S_{[2]}$ by $11\%$.
There is no Rscl-dependence for approximants (\ref{appexpl1}). 
The result $S(-Q^2)$ of the ECH
method is scheme- and RScl-independent, since this method is
local (i.e., a specific choice of RScl and scheme).

If changing the scheme more drastically, e.g.  
${\overline {\rm MS}}\!\mapsto$ `t Hooft scheme
($c_2\!=\!4.471\!\mapsto\!0$; $c_3\!=\!20.99\!\mapsto\!0$),
predictions for $S(-Q^2)$ of the global methods $[1/2]$,
$[2/2]^{1/2}$ and (\ref{appexpl1})
change by $1.2\%$ ($0.1273\!\mapsto\!0.1260$),
$1.2\%$ ($0.1326\!\mapsto\!0.1342$) and
$4.4\%$ ($0.1378\!\mapsto\!0.1.439$), respectively.
Scheme dependence (i.e., $c_2$-dependence) 
of (\ref{appexpl1}) does present
a problem in this case.

\subsection{A case of a Minkowskian observable}

\vspace{0.1cm}
 
\beq
R(p^2\!=\!s) = \frac{ {\sigma}(e^+e^- \to \gamma^{\ast} \to
{\rm hadr.})}{ {\sigma}(e^+e^- \to \mu^+ \mu^-)} .
\label{Rs}
\eeq
Here, $p^2$$=$$(p_e\!+\!p_{\bar e})^2$$=$$s$$>$$0$ is Minkowskian. 
Approximants generally lose 
predictability when applied to such observables
\cite{Kataevetal}. The associated
Adler function $D(p^2$$=$$-Q^2)$ ($Q^2$$>$$0$), 
RScl- and scheme-invariant, is Euclidean
\beq
D(-Q^2) = Q^2 \int_0^{\infty} ds R(s) (s+Q^2)^{-2} .
\label{D1}
\eeq
We apply approximants to it. The TPS up to NNLO
has been calculated for $D$ \cite{GKL}. For $n_f$$=$$5$, 
$q^2_{\rm RScl}\!=\!-Q^2$, in ${\overline {\rm MS}}$, it is
\beq
D = \frac{11}{3} [ 1 + d^{({\rm full})} ],
\quad d^{({\rm full})} = d + d^{(l.l.)},
\label{D2}
\eeq
\vspace{-0.5cm}
\bea
d(-Q^2)& =& a(-Q^2) [ 1 + 1.4092 a(-Q^2)
\nonumber\\
&&- 0.6812 a^2(-Q^2) + \cdots ],
\label{D3}
\\
d^{(l.l.)}(-Q^2) &=& a^3(-Q^2)(-0.3756) + \cdots.
\label{Dll}
\eea
We take $\sqrt{Q^2}\!=\!34$ GeV, $a(-Q^2)\!=\!0.0452$.
The $d^{(l.l.)}$ is from light-to-light diagrams, should not 
be included in the approximants since the resummations cannot
``see'' separately diagrams of fundamentally different
topologies. 
Reexpanding the approximants to $d(-Q^2)$ 
(\ref{D3}) in powers of $a(-Q^2)$, we predict
$d_3$ in series (\ref{D3}) (note: $d_1\!=\!1.4092$, 
$d_2\!=\!-0.6812$).
Then relation (\ref{D1}) allows us
to obtain the coefficients $r_j$ of the expansion
of $r(s)$ in powers of $a(-s)$
[$R\!=\!(11/3)(1\!+\!r^{({\rm full})})$,
$r^{({\rm full})}\!=\!r\!+\!r^{(l.l.)}$, 
$r^{(l.l.)}\!=\!d^{(l.l.)}$, $(-Q^2\!=\!-s)$]
\beq
r(s)=a(-s)[1 + r_1 a(-s) + \cdots],
\label{rs}
\eeq
with $r_1\!=\!d_1\!=\!1.4092$, $r_2\!=\!d_2\!-\!\pi^2 
\beta_0^2/3\!=\!-12.767$, $r_3\!=\!d_3\!-\!89.190$, etc. 
Thus, we can predict also the coefficient 
$r_3$ of $r(s)$. Results are given in Table 2. 
\begin{table*}[hbt]
\setlength{\tabcolsep}{0.4pc}
\catcode`?=\active \def?{\kern\digitwidth}
\caption{Predictions of various approximants for
the ratio $R(s)$ (with $n_f\!=\!5$) in the ${\overline {\rm MS}}$
($c_2\!=\!1.475$) and the 't Hooft ($c_2\!=\!0$) scheme 
(in brackets). $\sqrt{Q^2}\!=\!34$ GeV;
RScl chosen: $q^2_{\rm RScl}\!=\!-Q^2$; $a(-Q^2)\!=\!0.0452$.}
\label{tab2}
\begin{tabular*}{\textwidth}{@{}l@{\extracolsep{\fill}}rrrr}
\hline 
Approx. & $d(-Q^2)$ & $d_3^{\rm pr.}$ & 
$d_{[3]}(-Q^2,q^2_{\rm RScl})^{\rm pr.}$ & $r_3^{\rm pr.}$ \\
\hline
\hline
(TPS) & $0.048016$ ($0.047976$) & -- & $0.048016$ ($0.047976$) & -- \\
\hline
$[1/2]_d$ & $0.047996$ ($0.047974$) & $-4.72$ ($-0.56$) &
$0.047996$ ($0.047974$) & $-93.91$ ($-89.75$) \\
$[2/2]^{1/2}_{d^2}$ & $0.047978$ ($0.047965$) & $-8.48$ ($-2.24$) &
$0.047981$ ($0.047967$) & $-97.67$ ($-91.43$) \\
$\sqrt{A^{[2/2]}_{d^2}}$ & $0.047931$ ($0.047939$) &    
$-15.42$ ($-6.89$) & $0.047952$ ($0.047948$) & $-104.61$ ($-96.08$) \\
\hline
$d_{[2]}^{({\rm ECH})}$ & $0.047959$ ($0.047959$) & $-7.65$ ($-3.50$) & 
$0.047984$ ($0.047962$) & $-96.84$ ($-92.69$) \\
\hline
\end{tabular*}
\end{table*}
In brackets are results in the 't Hooft scheme
($c_j\!=\!0$ for $j\!\geq\!2$).
Predictions of the PA methods and the ECH are clustered.
Those of our method are
slightly, but significantly, detached from them.
Scheme ($c_2$-)dependence for the global approximants
is weak (cf.~last two digits in 
$2^{\rm nd}$ and $4^{\rm th}$ columns). 
Further, if the RScl is changed
$q^2_{\rm RScl}\!=\!-Q^2\!\mapsto\!-Q^2/4$,  
the $d(-Q^2)$'s of the PA's change about twice as strongly
as when the scheme is changed $c_2\!=\!1.475\!\mapsto\!0$.
Approximant (\ref{appexpl1}) does not change when RScl changes.

\section{Summary}
For a given $S_{[n]}$ (TPS) of
an observable $S$, we can construct an RScl-independent 
approximant which reproduces that TPS to the
given order $\sim$$a^{n+1}$. It is global, i.e.,
it goes beyond the (polynomial) form of the TPS and could thus
give us some clues to the nonperturbative effects --
possibly in contrast to the local methods (ECH, PMS).
It is, in principle, an improvement over
another global approximant -- the diagonal Pad\'e approximant
(dPA), the latter being RScl-invariant only in the large-$\beta_0$
limit. The question of how to eliminate the second major
source of unphysical dependence, the scheme ($c_2$-)dependence,
remains open. One possibility would be to choose an
``optimal'' $c_2$ (open problem). Another would be to
extend the method so as to give us, in a {\em global\/}
manner, approximants that are simultaneously RScl- and
$c_2$-independent (an open problem).

\end{document}